**Temperature dependence of the spontaneous magnetization of Ni2MnGa and other ferromagnets. The superellipse equation.**

Alexej Perevertov

**AFFILIATION**

FZU - Institute of Physics of the Czech Academy of Sciences, Department of Magnetic Measurements and Materials, 18200 Prague, Czech Republic

E-mail: perever@fzu.cz

## ABSTRACT

*The temperature dependence of the spontaneous magnetization of Ni2MnGa and other ferromagnets can be described in reduced coordinates by the superellipse equation using a single dimensionless parameter. This critical exponent parameter equals 2.4 for Ni2MnGa, 2.7 for nickel and cobalt, 3 for iron and 2.05 for gadolinium. Because reduced magnetization and reduced temperature enter the equation symmetrically, the $M_S(T)$ dependence can be measured only in the low-temperature range, from 0 to 0.5TC. The magnetization curve from $0.5T_C$ to $T_C$ can then be obtained by interchanging reduced magnetization and temperature in the superellipse equation. In this way, the experimentally challenging task of measuring spontaneous magnetization near $T_C$ is avoided, as the behavior near $T_C$ is effectively determined from measurements performed near $T = 0$. The $M_S(T)$ dependence in a whole temperature range is fully determined by two parameters – the Curie temperature, $T_C$ and the critical exponent, $\eta$.*

## I. INTRODUCTION

The spontaneous magnetization is the most important property of a ferromagnetic materials and is a quantity of fundamental significance. It is a result of local alignment of magnetic moments of neighboring atoms parallel to each other even in the absence of an external applied field. The mechanism for the appearance of spontaneous magnetization was proposed by P. Weiss in 1907 [1-5]. He assumed that aligned magnetic moments produce a large internal magnetic field that he called "the molecular field". Above the Curie temperature, $T_C$ the spontaneous magnetization, $M_S$ in a ferromagnetic material is lost and the material is considered to be paramagnetic [4].

The first theory of ferromagnetism by Weiss, which is used today, is based on the paramagnetism model of Langevin. Later it was extended to the more general Brillouin theory of localized magnetic moments [1-5]. In this model the spontaneous magnetization in the absence of the applied field is given by:

$$\frac{M_S}{M_0} = B_J\left(\frac{3J}{J+1}\frac{M_S}{M_0}\frac{T_C}{T}\right), \qquad (1)$$

where $J$ is the total angular momentum. $B_J$ – the Brillouin function.

This equation is of the $y = f(x,y)$ type and for arbitrary $J$ it can't be solved analytically. The $M_S(T)$ curve is usually obtained by a graphical method [1-5]. For $J = \infty$ the Brillouin function reduces to the Langevin function.

The experimental determination of temperature dependence of the spontaneous magnetization is a big challenge since $M_S$ can't be measured directly [6-17]. One can measure the sample magnetization (total magnetic moment) that can be small or even zero in the absence of the external magnetic field because a ferromagnetic material is subdivided into magnetic domains to minimize the magnetostatic energy.

To measure the spontaneous magnetization one has to apply the magnetic field large enough to align all magnetic moments parallel to the field thus creating a single-domain state.

The problem here is that the external field strongly influences the temperature dependence of the spontaneous magnetization near the Curie point – the Curie transition vanishes. One faces two mutually exclusive requirements – to apply zero field near the Curie point in order not to cancel the Curie transition and at the same time to apply large enough field to saturate the sample. It was proposed already in 1930s to measure a family of $M_S(T, H=const)$ curves at different applied fields, $H$ and then to find the zero-field curve, $M_S(T, 0)$ by the extrapolation to $H=0$ [2]. At low temperatures, there is usually no difficulty to measure $M_S$ – the magnetization changes very slowly with temperature ($dM_S/dT = 0$ at $T = 0$K) and the applied field has a very little effect on the curve [1-2]. However, change of magnetization at high temperatures close to $T_C$ becomes very rapid ($dM_S/dT = \infty$ at $T = T_C$), so that the measurement accuracy quickly deteriorates as the Curie point is approached.

## II. EXPERIMENTAL

In this work we use a sample-yoke method to reduce the field necessary to saturate the sample [18]. A 60mm-long Ni2MnGa single-crystal sample is measured along the magnetically easy direction – [100] in the austenitic state (T > 210K). In this case the sample can be saturated by a very low field of 1600 A/m ($\mu_0 H = 2$ mT). So, the field needed to saturate the Ni2MnGa single-crystal sample is 1000 times lower compared to commonly used VSM measurements [19] and the $M_S(T)$ is nearly unaffected near $T_C$. Similar to phase transformations in cobalt the martensite to austenite phase transformation in Ni2MnGa at 220 K does not affect significantly $M_S$.

In Fig. 1 the spontaneous magnetization, $M_S$ as a function of temperature, $T$ for a Ni2MnGa single crystal is shown. Results of sample-yoke measurements above the martensitic transformation, $T_M = 210$ K are shown for the applied field, $H = 5000$ A/m ($\mu_0 H = 6$ mT). Above $T_M$ Ni2MnGa is in the austenitic state with a cubic crystal structure. Below $T_M$ Ni2MnGa is in the martensitic state with a monoclinic crystal structure. The sample is divided into regions with three different crystal orientations – twins [20]. Since the magnetocrystalline anisotropy of Ni2MnGa martensite is very high [20], the sample cannot be saturated in the sample-yoke configuration – the field required is much larger than that produced by the driving coil on the yoke. To obtain the spontaneous magnetization below $T_M$ a small sample was measured in VSM at the applied field, $H = 1.6$ MA/m ($\mu_0 H = 2$ T), which is large enough to overcome the magnetocrystalline and shape anisotropies to saturate the sample. The spontaneous magnetization at $T = 0$, $M_0 = 0.816$ MA/m ($\mu_0 M_0 = 1.02$ T). The Curie temperature, $T_C$ from the sample-yoke measurement is 371 K. At the martensite-to-austenite phase transformation $M_S$ changes insignificantly – by 3.5% only. Small dip in the magnetization at the premartensitic transformation at $T_{PM} = 264$ K is even smaller – about 1.5% [19]. Between 210 K and 355 K (from $0.57T_C$ to $0.95T_C$) results of both measurements are identical. Above 355 K ($0.95T_C$) the VSM $M_S(T)$ curve is distorted due to large applied field of 1.6 MA/m (2 T) – the Curie transition vanishes [2,5,21-22]. The magnetization does not go to zero at this field even at 800 K. So, to study the $M_S(T)$ curve in the full temperature range from 0 to $T_C$ we use a combination of two measurements: VSM from 0 to $0.57T_C$, both from $0.57T_C$ to $0.95T_C$ and the sample-yoke close to $T_C$.

## III. RESULTS

In Fig. 2 $M_S(T)$ fits using Eq. (1) for $J = \infty$ and $J = ½$ are shown together with experimental data. The $J = ½$ curve agrees with experimental data at lower temperatures, but disagrees with experiment above $0.95T_C$ and gives a different value for $T_C$ – 358 K instead of 371 K. The $M_S(T)$ curve for $J = \infty$ agrees with experiment only near $T_C$ and completely fails at other temperatures.

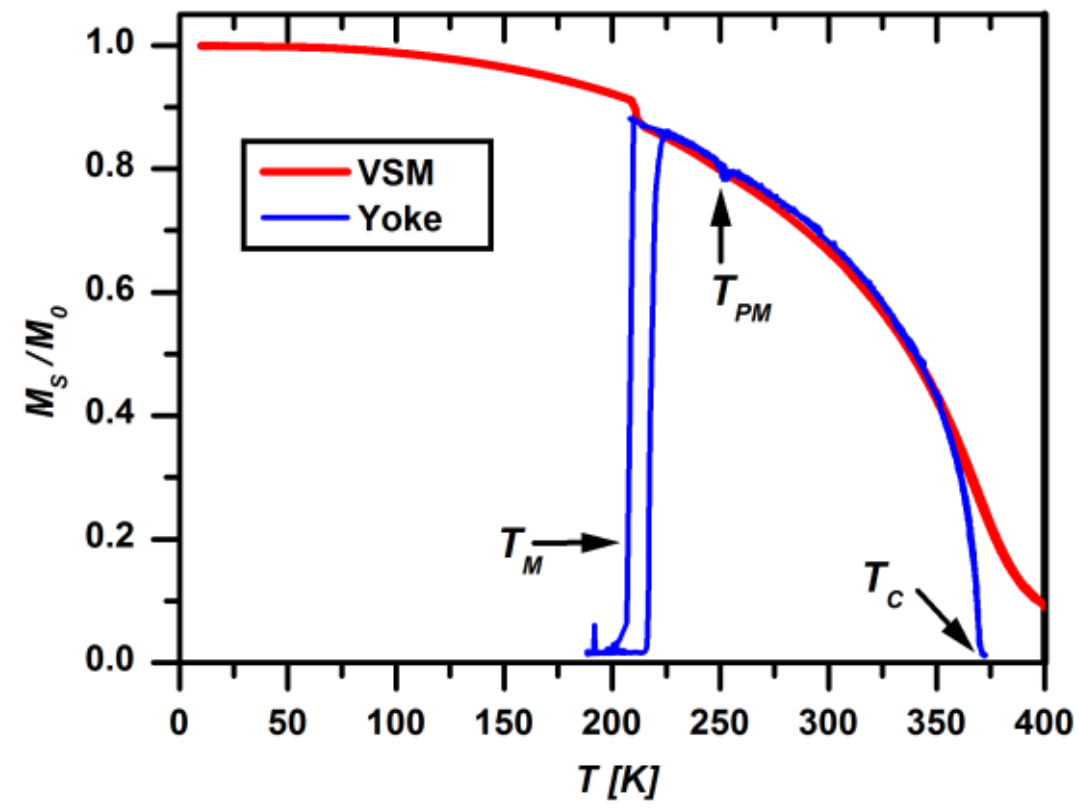

FIG. 1 Spontaneous magnetization, $M_S$ as a function of temperature, $T$ for sample-yoke measurements ($T > T_M$) at the applied field, $H$ = 5000 A/m and measurements in VSM at the applied field, $H$ = 1.6 M A/m from 10 to 400 K. $M_S$ at $T$ = 0, $M_0$ = 0.816 MA/m ($\mu_0 M_0$ = 1.02 T). The Curie temperature, $T_C$ from the sample-yoke measurement is 371 K.

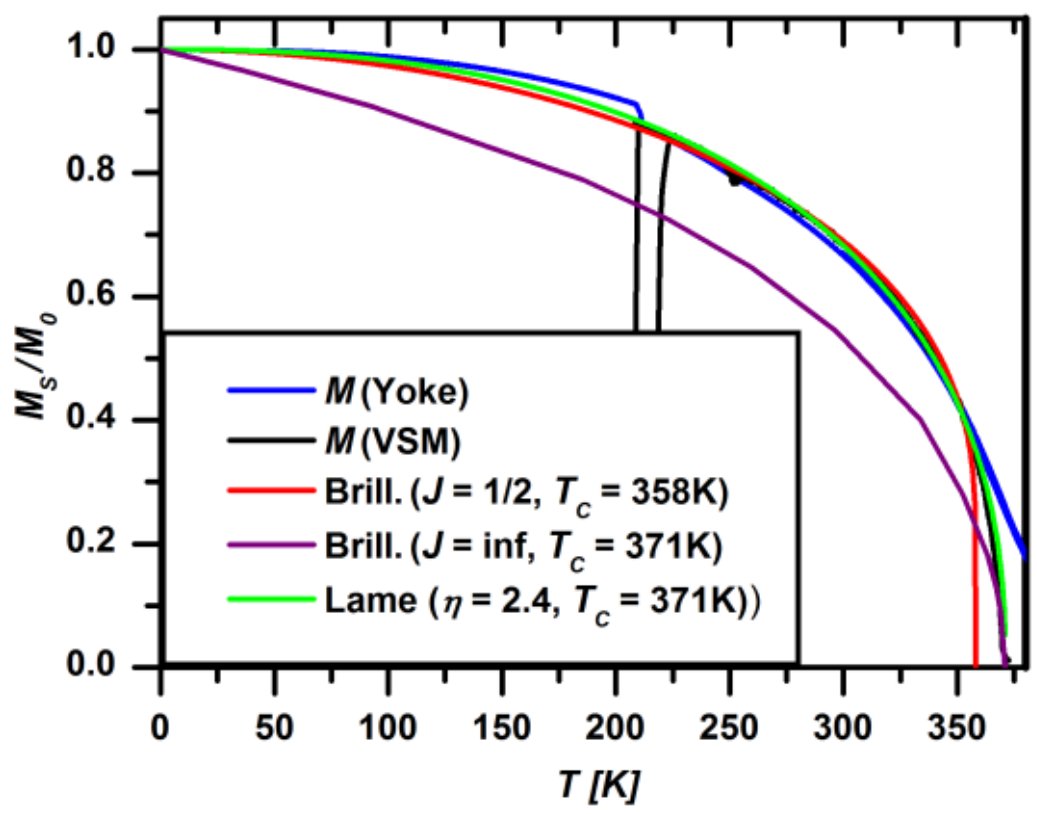

FIG. 2. Fits of experimental data by classical equation the Brillouin theory (see Eq. (1)) for $J$ = ½ and ∞ together with the fit by the Lamé curve ($\eta$ = 2.4).

## IV. DISCUSSION

In 2005 a phenomenological $M_S(T)$ relation was introduced by Kuz'min [23] to approximate the $M_S(T)$ curve at all temperature range from o to $T_C$ in the following form:

$$m(\tau) = \left[1 - s\tau^{\frac{3}{2}} - (1-s)\tau^{p}\right]^{1/3}, \tag{2}$$

where $s$ and $p$ are parameters, p > 3/2, s > 0, $m = M_S/M_0$ is the reduced spontaneous magnetization at a reduced temperature $\tau = T/T_C$.

This equation was constructed to obey the classical power laws at low temperatures and around $T_C$. It showed good agreement with a number of ferromagnetic materials and is used by many authors. At present, researches use the classical Brillouin-Langevin theory or/and the empirical equation by Kuz'min [23-29]. In this work we want to advance the field by introducing a much simpler $M_S(T)$ relation.

We propose the magnetization and temperature relation in the form of the superellipse equation or the Lamé curve:

$$(M_S/M_0)^{\eta} + (T/T_C)^{\eta} = m^{\eta} + \tau^{\eta} = 1, \tag{3}$$

where $m$ is the reduced magnetization, $\tau$– the reduced temperature and $\eta$ is the parameter (critical exponent).

Applying this relation to our data on Ni2MnGa single crystal, we see that it gives very good agreement with the experimental data in a whole temperature range for $\eta = 2.4$ (see Fig. 2).

We also applied the Eq. (3) to experimental data by other authors on four main ferromagnetic elements: iron, nickel, cobalt and gadolinium (see Fig. 3) [30-35]. The agreement with experimental data was very good. The power coefficient, $\eta$ is 2.7 for nickel and cobalt, 3 for iron and 2.05 for gadolinium. It is interesting that the magnetization curve for cobalt and nickel is nearly identical when plotted in the reduced coordinates.

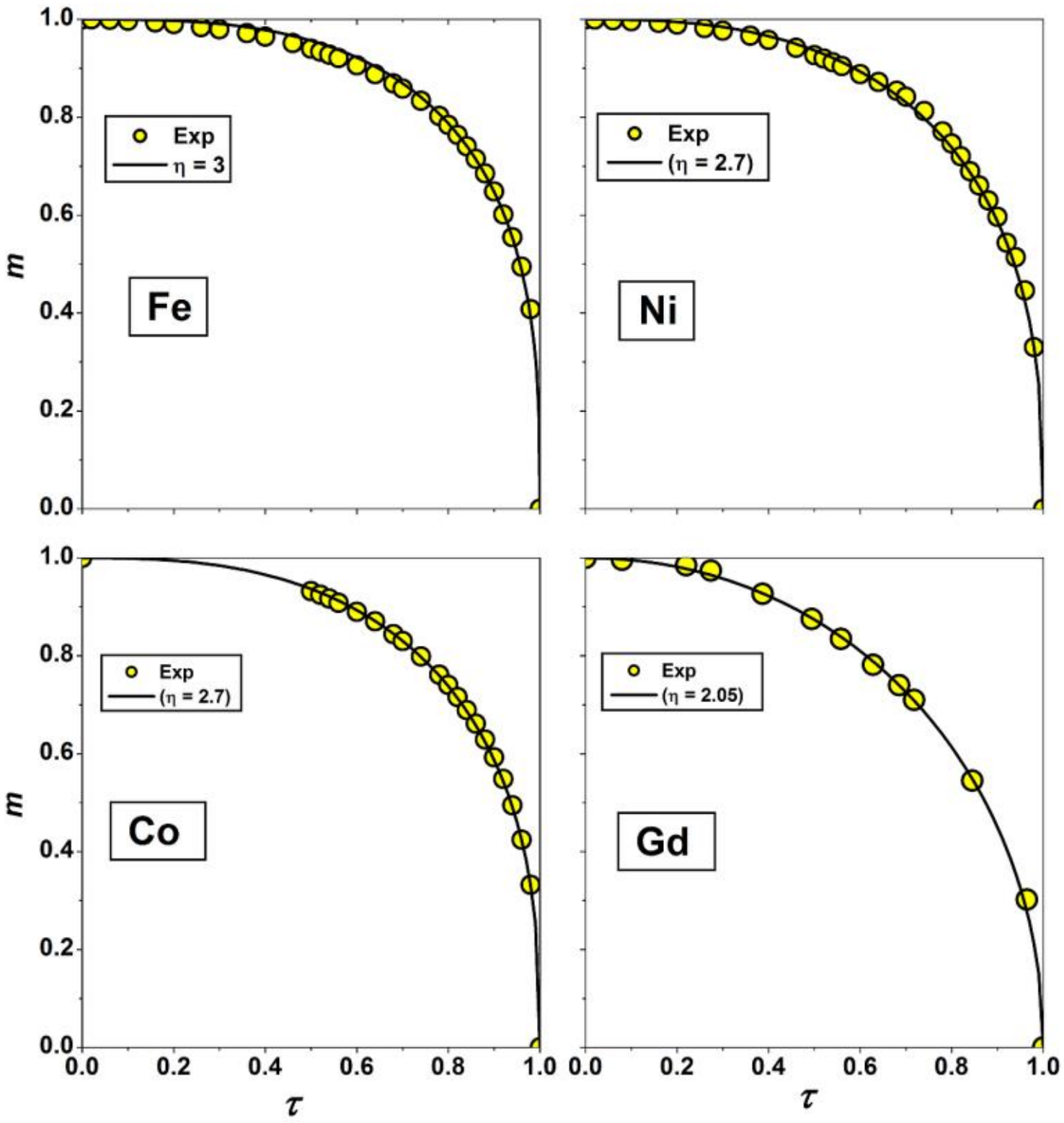

FIG. 3. Reduced magnetization vs. reduced temperature for Ni, Co, Fe (data from Ref.30) and Gd (Ref.34) and corresponding fits by the superellipse Lamé function with $\eta = 2.7$ for Ni and Co, 3 for Fe, 2.05 for Gd.

Close to the Curie temperature the magnetization from Eq. (3) can be approximated as:

$$(M/M_0)^{\eta} = 1 - ((T-T_C)/T_C + 1)^{\eta} = 1 - (1 + \eta\,(T-T_C)/T_C + 0.5\,\eta(\eta-1)\,[(T-T_C)/T_C]^2 + \ldots) =$$

$$= \eta\,(T_C-T)/T_C - 0.5\,\eta(\eta-1)\,[(T_C-T)/T_C]^2 - \ldots \sim \eta\,(T_C-T)/T_C \qquad (4)$$

$$m = (\eta\,(1-\tau))^{1/\eta} \qquad (5)$$

The Eq. (5) gives the critical exponent for the spontaneous magnetization $\beta = 1/\eta$. The expansion in Eq. (4) is the classical binomial expansion $(1+x)^{\eta}$ with $x = (T_C-T)/T_C$. The logical question arises the same as in the classical mean field theory – at which condition only the first two terms can be left, $1 + \eta x$. $x$ has to be as small as possible, preferably less than 0.001. From other side, the minimum experimental error in $T_C$ determination is also around 0.1%. As we discussed above, the spontaneous magnetization can't be measured directly and has the highest relative error namely close $T_C$. At $0.9T_C$ ($x$=0.1) $m \sim 0.6$ for Ni, Co and Fe, which is too high. $m \sim 0.2$ for $x$=0.01. Our analysis of $m(\tau)$ data for all materials present in literature showed that $1.4<\eta<3$. Let us consider two extreme cases: $x$=0.05, $\eta$ = 1.4 and 3.

Leaving first three terms in the binomial expansion gives for $\eta$ = 1.4: $m^{\eta} \sim 1.4x + 0.28x^2$. For $x$ = 0.05 the third quadratic term in the expansion is 100 times lower than the second linear term and can be safely neglected. For $\eta$ = 3: $m^{\eta} \sim 3x + 3x^2$. For $x$ = 0.05 the quadratic term is 20 times lower and can be also neglected. Lamé functions with their approximation by Eq. (5) are shown in Fig. 4.

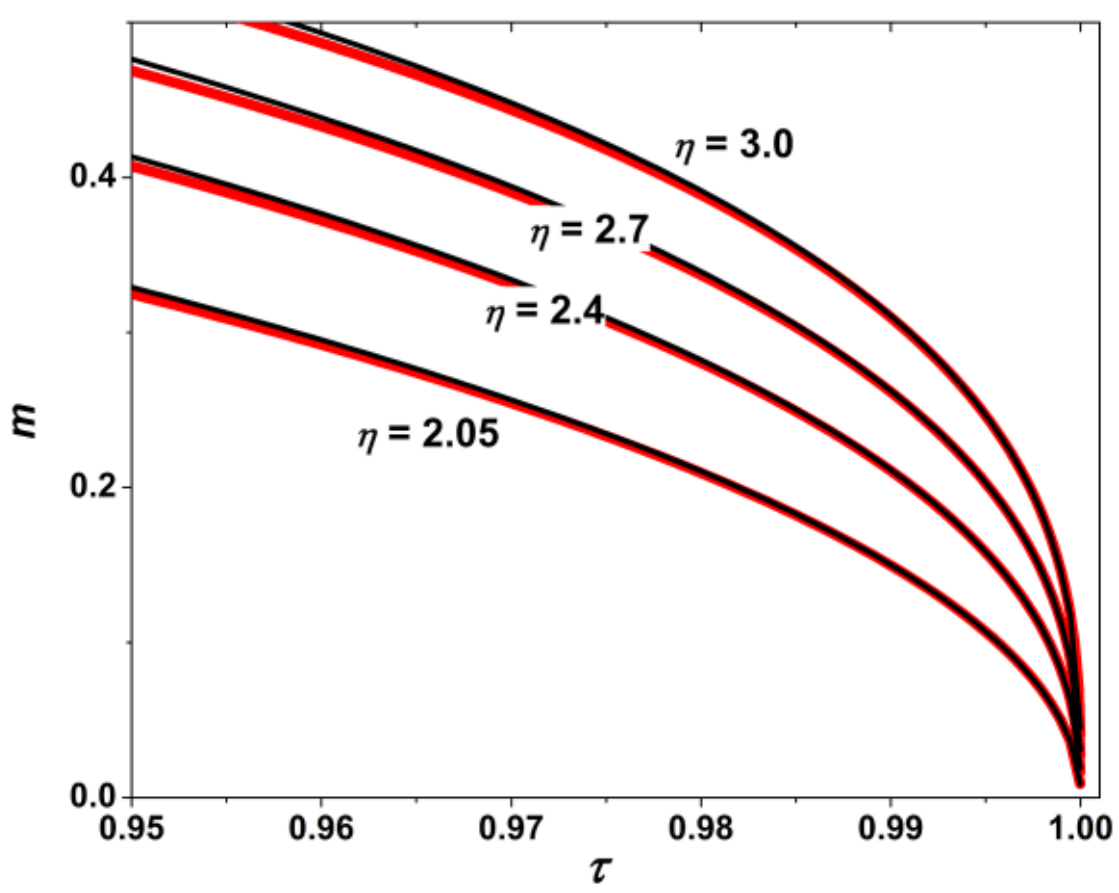

FIG. 4. Lamé functions with their approximation by Eq. (5) close to $T_C$ for different $\eta$.

The critical exponent $\beta$ derived from Lamé functions and from classical models: mean field, Ising and Heisenberg [1-5] are given in the Table 1.

Table 1. The critical exponent $\beta$ derived from Lamé functions and from classical models.

| | $\eta$ = 2.05 | $\eta$ = 2.4 | $\eta$ = 2.7 | $\eta$ = 3 | MF | Ising | *H* |
|---|---|---|---|---|---|---|---|
| $\beta$ | 0.488 | 0.417 | 0.370 | 0.333 | 0.5 | 0.326 | 0.365 |

At very low temperatures the magnetization from Eq. (3) is the same as for all temperatures:

$$m = (1 - \tau^{\eta})^{1/\eta} \tag{6}$$

In the Brillouin theory, $M_S \sim (T_C\text{-}T)^{1/2}$ [1-5].

As in the Brillouin theory, the magnetization approaches zero temperature with a horizontal slope ($dM_S/dT = 0$), as required by thermodynamics [3]. The magnetization and temperature are equivalent in Eq. (3), so the magnetization approaches the Curie temperature, $T_C$ with a vertical slope ($dM_S/dT = \infty$). This makes the experimental determination of the $M_S(T)$ curve near $T_C$ being an extremely difficult task.

So, it is hard to say which scaling law is valid for the spontaneous magnetization close to $T_C$ because of large experimental error in this temperature region. The Curie-Weiss law is based on the assumption that the magnetization is given by a Brilluin function, which in fact fails to predict accurately experimental $m(\tau)$ curves of ferromagnetic materials.

From Eq. (3) we see that plotting $m^{\eta}$ as a function of $\tau^{\eta}$ gives the straight line $y = 1 - x$ for a correct value of $\eta$. In Fig. 5a such plots for experimental data on Ni2MnGa, nickel, cobalt, iron and gadolinium are shown. So, the simplest procedure to evaluate the power coefficient $\eta$ is to plot $m^{\eta}$ vs. $\tau^{\eta}$ for different values of $\eta$ (see Fig. 5b). Then by plotting the mean deviation of $m^{\eta}(\tau^{\eta})$ from $1 - x$ line for different $\eta$ one can find a best fit value, for which the mean deviation is zero (see Fig.6a). This method gives $\eta$ = 2.05 for Gd, 2.4 for Ni2MnGa, 2.67 for Ni and Co, and 2.9 for Fe. The error of the power coefficient estimation by this method is around ±0.05, which is mostly given by the accuracy of the experimental data. The direct fit of $m(\tau)$ by Lamé functions gives the same power coefficient for Gd and Ni2MnGa, but slightly higher values for Ni and Co (2.7) and 3.0 for iron (see Fig.6b).

The dependence of the spontaneous magnetization on the temperature in the form of the superellipse equation (See Eq. (3)) greatly simplifies the analysis and determination of experimental data. First of all, the $m(\tau)$ function is a very simple analytical function with only one unknown parameter – the critical exponent $\eta$. In the superellipse equation the arguments can be interchanged. One can notice that for most ferromagnets indeed the $m(\tau) = \tau(m)$ – the magnetization and temperature can be swapped in the temperature dependence [30-35]. So one can measure $M_S(T)$ from 0 to $0.5T_C$ and mirror the curve swapping $m$ and $\tau$ and thus obtaining $M_S(T)$ from $0.5T_C$ to $T_C$. As we showed above, the applied field of 1.6 MA/m had no effect on the $M_S(T)$ curve from 0 to $0.9T_C$, so one can safely measure the saturation magnetization in this range and obtain the rest of the curve from the corresponding part at lower temperatures. The Curie temperature can be obtained by other measurements such as a small-field susceptibility [13,19].

It is surprising that the symmetry of the dependence of the reduced magnetization on the reduced temperature has not been noticed until now (See Fig. 3). In contrast to the superellipse equation, the $m(\tau)$ in the classical theory (see Eq. (1)) is not symmetrical and can't be mirrored (see Fig. 7).

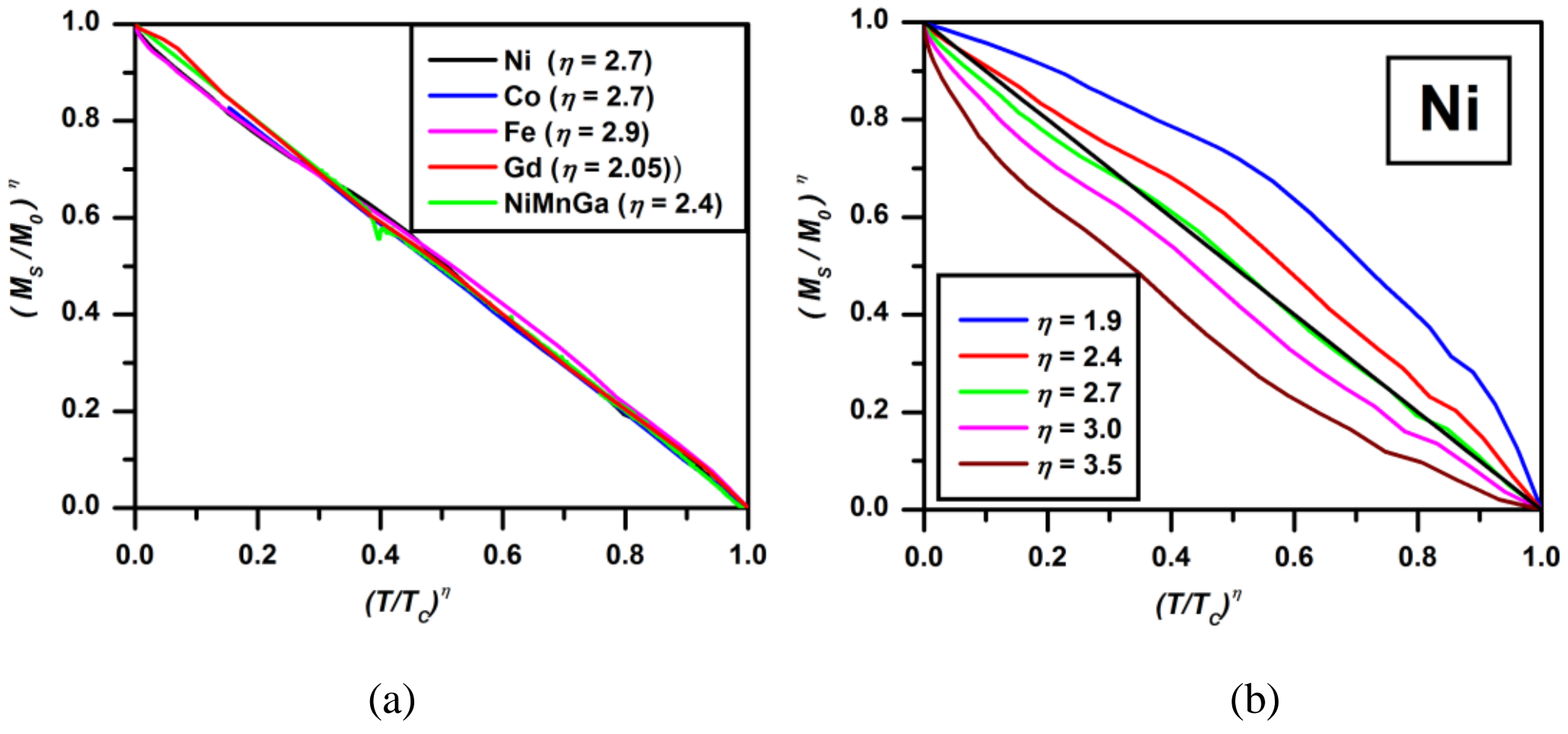

FIG. 5. $m^{\eta}(\tau^{\eta})$ plots for nickel, cobalt, iron, gadolinium and NiMnGa (a) and $m^{\eta}(\tau^{\eta})$ plots for nickel for different values of $\eta$ (b).

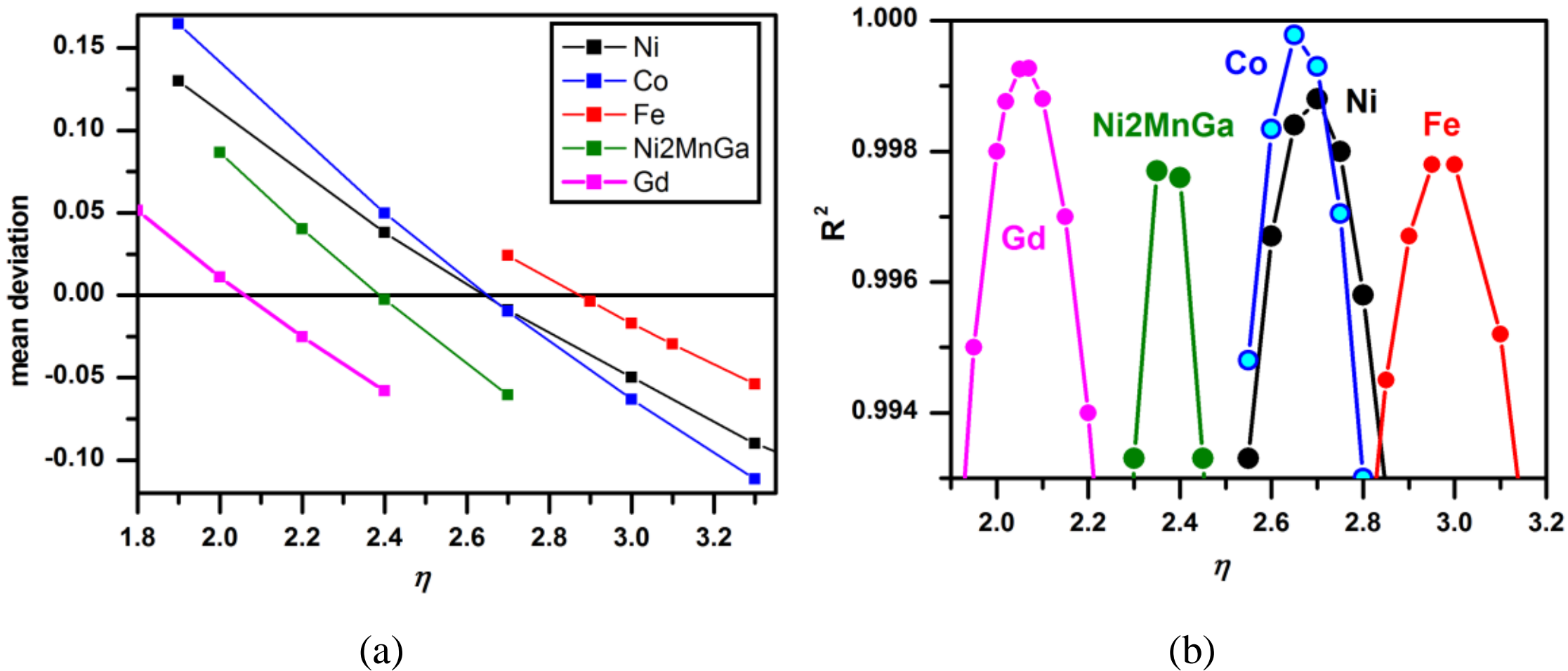

FIG. 6. Mean deviation of $m^{\eta}(\tau^{\eta})$ plots from $1 - x$ line (a) and the coefficient of determination $R^2$ (b) *of* $m(\tau)$ plots fitting for different $\eta$ for Ni, Co, Fe, Gd and Ni2MnGa.

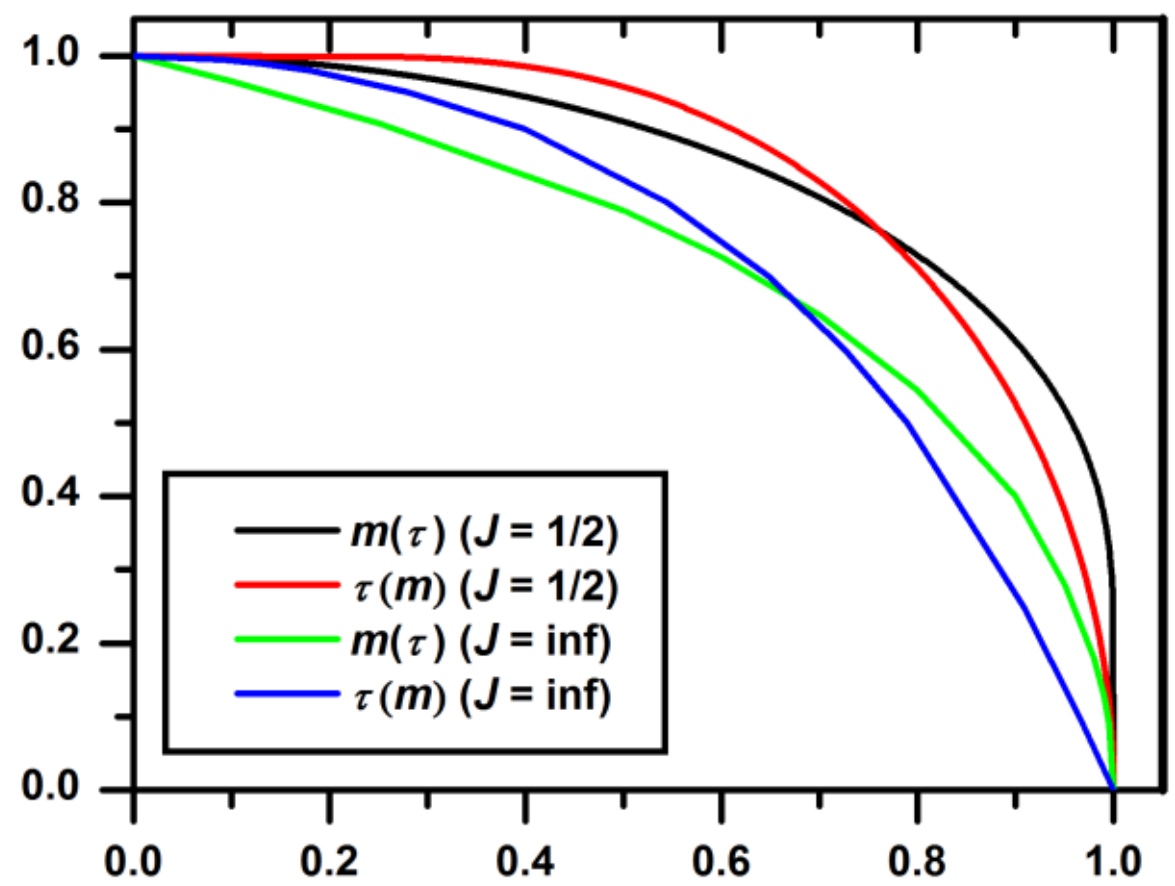

FIG. 7. $m(\tau)$ and the mirrored $\tau(m)$ dependences by Eq. (1) for $J = ½$ and ∞.

There were a number of attempts to approximate $M_S(T)$ dependence by simple analytical functions instead of classical ones by the Brillouin theory. In 1972 the ellipse equation ($\eta = 2$) was proposed to approximate the $M_S(T)$ of iron, nickel and invar at low temperatures below $0.45T_C$ [36-37]. At higher temperatures this formula failed to describe the experimental data. As we showed above, the reason is a wrong power coefficient $a = 2$, that should be 2.7 for nickel and cobalt and 3 for iron. The ellipse equation also appears in the model of weak itinerant ferromagnetism [38-39].

The problem in previous studies researches tried to adapt their equations to the classical power laws near $T = 0$ and $T_C$, which automatically excludes symmetry in the $m(\tau)$ equation [23-29, 36-40]. In these analytical relations only whole numbers or fractions that are multiples of 1/2 and 1/3 were used in the power coefficients. In our work we showed that they are irrational. Due to these restrictions, the superellipse equation have never been proposed. In 2015 Evans et al. proposed the simplest form of an equation as an interpolation of the Bloch law and critical behavior given by the Curie-Bloch equation [40]:

$$m = (1 - \tau^a)^b,\ b = 1/3 \tag{7}$$

The restrictions by the classical laws for $T$=0 and $T$=$T_C$ forced the authors to fix the power exponent, $b$ = 1/3 and did not allow them to come to a much simpler symmetrical superellipse equation by making $b$ irrational and equal to $1/a$. As we discussed above, the accuracy of the experimental values of $M_S$ decreases exponentially approaching $T_C$. So it is difficult to check experimentally which power law the $M_S(T)$ curve obeys. We can only say that the curve approaches $T_C$ with a vertical slope. Also our analysis of the experimental date (See Fig. 5) allows us to assume that the $M_S(T)$ curve has the superellipse shape also in the vicinity of $T_C$.

We applied the Eq. (3) to about forty different materials and it agreed very well with experimental data [41]. The $m(\tau)$ curve was symmetrical in most cases, so we did not need to modify the superellipse equation. From other side, the Brillouin functions are asymmetrical. They can be approximated with a good accuracy nearly in all temperature range by the modified superellipse equation (see Fig. 8 and Table 2)

$$m^{\mathrm{a}} + \tau^b = 1, \tag{8}$$

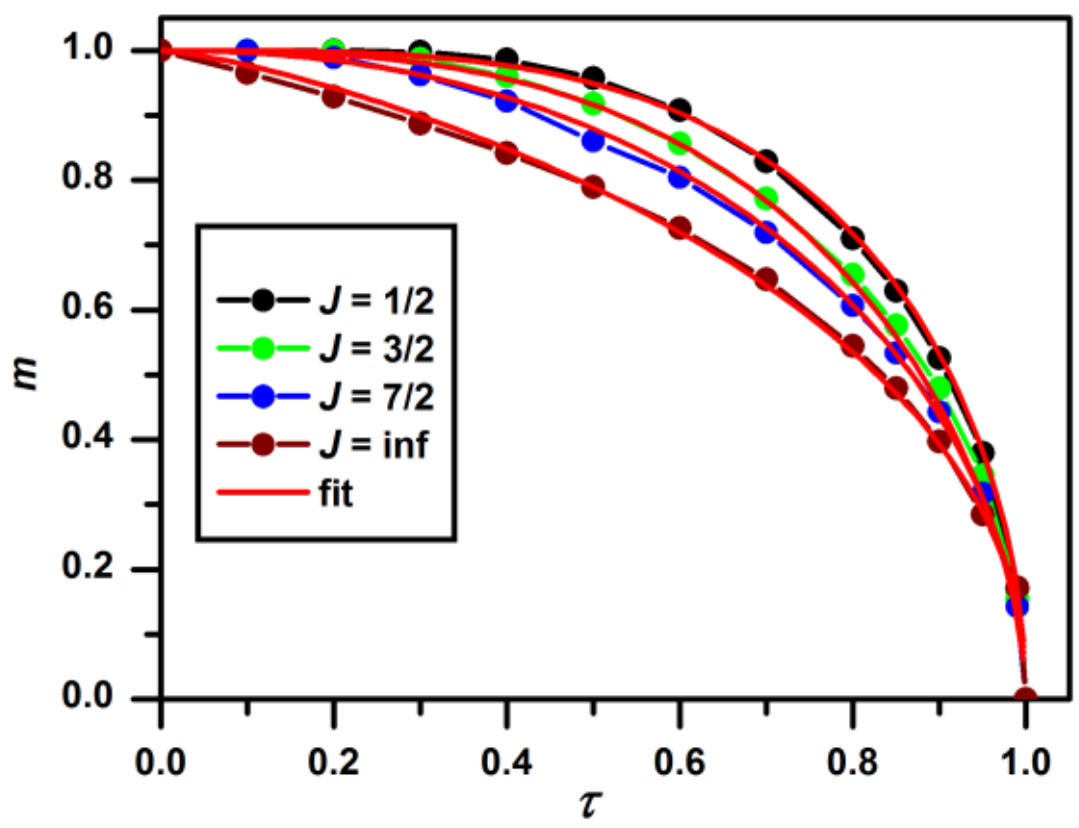

Fig. 8. Fits of Brilluin functions by Eq. (8).

Table 2. Fitting confidents of Brilluin functions fits by Eq. (8) for different *J*.

| | $J = 1/2$ | $J = 1$ | $J = 3/2$ | $J = 2$ | $J = 5/2$ | $J = 7/2$ | $J = \infty$ |
|---|---|---|---|---|---|---|---|
| *a* | 1.9 | 1.7 | 1.7 | 1.8 | 1.8 | 1.9 | 1.3 |
| *b* | 3.4 | 3.2 | 2.85 | 2.6 | 2.5 | 2.2 | 2.2 |

Previously we have shown that for the magnetization curve, *M(H)* of polycrystal ferromagnets also a simple analytical function can be used – the arctangent instead of the complex Langevin equation [42]. Here we showed for the $M_S(T)$dependence one can use a very simple analytical formula – the superellipse equation. In both cases these simple analytical expressions give better agreement with experimental data comparing to much more complex classical equations, that can't be solved analytically. It greatly simplifies analysis of *M(H)* and $M_S(T)$ dependences by using the arctangent and Lamé functions correspondingly.

**SUMMARY.** In summary, in this work we experimentally studied the temperature dependence of the spontaneous magnetization of Ni2MnGa. The relation between the magnetization and temperature for this material and several other ferromagnets (nickel, iron and cobalt) is described by the superellipse equation with a single dimensionless parameter in the reduced coordinates. The critical exponent parameter, $\eta$ is 2.4 for Ni2MnGa, 2.7 for nickel and cobalt, 3 for iron, and 2.05 for gadolinium. The equivalence of the reduced magnetization and temperature in the superellipse equation ($m(\tau) = \tau(m)$) and in the experimental data is very intriguing. Due to this symmetry, one can measure the $M_S(T)$ dependence of a material at lower temperatures from, say 0 to $0.5T_C$ and then inverse $M_S(T)$ dependence to obtain the curve from $0.5T_C$ to $T_C$ by swapping the reduced magnetization and temperature in the equation. In such a way we eliminate a very difficult task of the spontaneous magnetization measurements near $T_C$ by studying $M_S(T)$ around $T = 0$ instead. The dependence of the critical exponent parameter, $\eta$ on a material crystal and electronic structure is an interesting topic for investigation.

## DATA AVAILABILITY

The data that support the findings of this study are available from the corresponding authors upon request. The data are also available in Zenodo at https://doi.org/ 10.5281/zenodo.18132106 (Ref. 43).

## ACKNOWLEDGMENTS

The authors acknowledge the assistance provided by the Ferroic Multifunctionalities project, supported by the Ministry of Education, Youth, and Sports of the Czech Republic. Project No. CZ.02.01.01/00/22_008/0004591, co-funded by the European Union. The single crystal preparation was partially performed in MGML (http://mgml.eu/), which was also supported within the program of Czech Research Infrastructures (project no. LM2023065). We also appreciate partial financial support by the Czech Science Foundation (project no. 23-4806S),